\documentstyle[aps,prl,multicol,epsf]{revtex}
\epsfclipon
\begin{document}
\draft
\title{Anomalous finite-size effect in 
superconducting Josephson junction arrays
}
\author{Qing-Hu Chen$^{1,2}$, Lei-Han Tang$^{1,3}$,
and Peiqing Tong$^{1,4}$
}
\address{
$^{1}$Department of Physics, Hong Kong Baptist University, 
Kowloon Tong, Hong Kong\\
$^{2}$Department of Physics, Zhejiang University, 
Hang Zhou 310027, P. R. China\\
$^{3}$Institute for Theoretical Physics, University of California,
Santa Barbara, CA 93106\\
$^{4}$Department of Physics, Nanjing Normal University, 
Nanjing 210097, Jiangsu, P. R. China
}
\date{\today}
\maketitle

\begin{abstract}
We report large-scale simulations of the resistively-shunted
Josephson junction array in strip geometry.
As the strip width increases, the voltage first decreases following
the dynamic scaling ansatz proposed by Minnhagen {\it et al.}
[Phys. Rev. Lett. {\bf 74}, 3672 (1995)], and
then rises towards the asymptotic value predicted by
Ambegaokar {\it et al.} [Phys. Rev. Lett. {\bf 40}, 783 (1978)].
The nonmonotonic size-dependence is attributed to shortened life time
of free vortices in narrow strips,
and points to the danger of single-scale
analysis applied to a charge-neutral superfluid state.

\end{abstract}

\begin{multicols}{2}
Dissipation of the supercurrent through superfluid and superconducting
films is caused by the flow of free vortices transverse to the 
current\cite{feynman,ktb,blatt}.
This phenomenon was analyzed in detail by Ambegaokar, Halperin, 
Nelson, and Siggia (AHNS)\cite{ahns}.
In the context of a uniform superconducting film
in zero magnetic field, their theory predicts a power-law 
current-voltage ($I$-$V$) relationship with a temperature-dependent 
exponent. Despite the relatively simple construct of the theory,
its experimental and numerical verification has remained controversial
\cite{repaci,herb,mon,simkin,hwang,kim}. It has been noted that
boundary and finite-size effects can dominate the measured 
voltage drop across the sample at sufficiently low temperatures,
making it difficult to achieve an unambiguous comparison
between theory and experiments.

An alternative theory of vortex flow dissipation which also yields 
nonlinear $I$-$V$ curves in the superconducting state derives from 
the dynamic scaling hypothesis\cite{ffh,minn95}.
For superconducting films and Josephson junction arrays (JJA), 
the two approaches
differ in their predictions of the exponent $a(T)$ characterizing the
power-law behavior $V\sim I^a$ below the Kosterlitz-Thouless-Berezinskii 
(KTB) transition temperature $T_{\rm KT}$.
According to the AHNS theory,
\begin{equation}
a_{\rm AHNS}(T)=\chi+1,
\label{a_ahns}
\end{equation}
where $\chi=\pi J_R/k_BT$ and $J_R$ is the renormalized spin-wave stiffness.
On the other hand, the dynamic scaling analysis of
Minnhagen {\it et al.}\cite{minn95} suggests
\begin{equation}
a_{\rm MWJO}(T)=2\chi-1.
\label{a_mwjo}
\end{equation}
The two expressions coincide at $T_{\rm KT}$ where $\chi=2$, 
but the difference grows rapidly as one moves to low
temperatures. While there seem to be ample numerical support for the 
Minnhagen {\it et al.} scaling\cite{kim,minn95,weber,choi}, 
agreement with the AHNS theory has also been reported\cite{simkin}.
To rationalize the two scenarios, Bormann\cite{bormann} made an 
interesting suggestion
that strong current creates a dense set of vortex-antivortex
pairs, thereby invalidating the AHNS treatment.
Nevertheless, the AHNS theory should still apply at sufficiently weak
currents. This picture has not been borne out by a recent
numerical study\cite{kim} which shows persistently larger value of $a$ than
Eq. (\ref{a_ahns}) predicts.

In this Letter, we present $I$-$V$ results from large-scale simulations 
of the resistively shunted Josephson-junction (RSJ) array. 
A rectangular strip geometry is used to decouple boundary
effects introduced by the current leads from finite-size effects
arising from vortex/antivortex motion transverse to the current $I$.
The main finding of our work is the existence of
three distinct regimes as the strip width $L_y$ is varied.
For $L_y<l_{\rm b}\sim I^{-1}$, one is in the 
linear-response regime where the near-equilibrium dynamic scaling 
hypothesis applies. In this regime,
the voltage $V$ decreases as $L_y$ increases,
and reaches a minimum value predicted by Minnhagen {\it et al.}
When $L_y$ exceeds a second length scale $l_{\rm r}\sim I^{-\chi/2}$, 
$V$ saturates to the asymptotic value predicted by the AHNS theory.
The intermediate regime $l_{\rm b}<L_y<l_{\rm r}$ is characterized
by a {\it rising} $V$ as $L_y$ increases. This behavior is explained
in terms of ``self-recombination'' of vortex-antivortex pairs
under periodic boundary conditions. 

We begin with the model Hamiltonian of a two-dimensional (2D) 
JJA in zero magnetic field,
\begin{equation}
H =-J\sum_{\langle ij\rangle} \cos(\phi_i-\phi_j).
\label{Hamiltonian}
\end{equation}
Here $\phi_i$ denotes the phase of the superconducting order parameter on 
grain $i$, and $J$ sets the strength of the Josephson coupling between 
neighbouring superconducting grains.
The summation in Eq. (\ref{Hamiltonian}) is over all nearest neighbor
grain pairs. In the RSJ dynamics, the network is insulated from the 
substrate, but normal currents are allowed between neighboring grains,
in addition to the supercurrent.
The dynamics of the phase variables follows 
the Josephson relations plus the Kirchoff law for current conservation
at each node\cite{mon},
\begin{equation}
{\sigma\hbar \over 2e}\sum_{j\in{\rm n.n.}\ {\rm of}\ i}
\Bigl({d\phi_i\over dt}-{d\phi_j\over dt}\Bigr)
=-{2e\over\hbar}{\partial H\over\partial\phi_i}+I_{\rm ext}.
\label{dyn_eqn}
\end{equation}
Here $\sigma$ is the conductivity of normal links, and
$I_{\rm ext}$ is the external current source which is nonvanishing
only at the boundaries of the network. To model the effect of 
temperature $T$, 
a gaussian noise current $\eta_{ij}$ is added to each link between 
two adjacent nodes $i$ and $j$, with
$\langle \eta_{ij}(t)\eta_{i'j'}(t')\rangle=2\sigma k_BT
\delta_{ij,i'j'}\delta(t-t')$.

To integrate the coupled equations (\ref{dyn_eqn}), one needs to
first solve for $V_i=(\hbar/2e)d\phi_i/dt$. This can be done
efficiently using a pseudo-spectral algorithm.
In our simulations, the network is chosen to be a strip
of $L_x$ columns, with $L_y$ nodes in each column. 
Open boundary conditions are used in the $x$-direction (along
the strip) to allow for current inputs and outputs.
Periodic boundary conditions (PBC) are used in the $y$-direction
perpendicular to the strip. 
After performing the Fourier transform in the $y$-direction,
the left-hand-side of Eq. (\ref{dyn_eqn}) reduces to a tridiagonal 
form which can be easily solved by the Gauss elimination method.
The time stepping is done 
using a second-order Runge-Kutta method with $\Delta t=0.1$ (in units of
$\sigma\hbar/2eI_c$, where $I_c=2eJ/\hbar$ is the critical current through
each link)\cite{schein}. To minimize the boundary effects,
voltage drop across the sample is measured in the interior of the 
network. For the data presented below, we typically choose 
$L_x=2048+2\times 128$
and discard 128 columns at each end for this purpose.

When the external current $I$ into each boundary node is set to zero, 
we recover a dynamical version of the XY model whose equilibrium 
properties have been well characterized. In particular, the 
KTB transition temperature has been determined to be 
$T_{\rm KT}\simeq 0.89J/k_B$\cite{olsson}.
We have confirmed this value in the simulation by examining the
correlation function of the superconducting order parameter
and the helicity modulus (or $J_R$), which 
exhibit universal behavior at the transition. 
Direct measurement of the $I$-$V$ curve at this temperature has 
also confirmed the predicted exponent $a(T_{\rm KT})=3$. 

We now describe the main results of our simulation.
Figures 1(a) and 1(b) present the $I$-$V$ data
at $T=0.8J/k_B$ and $T=0.7J/k_B$, respectively, 
for four different array sizes: $L_y=8, 32, 128$, and 512.
Here the current $I$ is measured in units of $I_c$, and
$V$, the voltage drop per column, in units of $\hbar/2e$.
As the current $I$ decreases, the finite size effect
becomes more and more evident. Examining the data at
the smallest $I$ shown, we see a clear nonmonotonic size
dependence: for small values of $L_y$, $V$
decreases with increasing $L_y$. However, this trend is reversed
as $L_y$ is increased further. The solid and dashed lines in the
figure correspond to the power-law behavior predicted by AHNS
[Eq. (\ref{a_ahns})] and Minnhagen {\it et al.} [Eq. (\ref{a_mwjo})], 
respectively, using $\chi=2.8$ at $T=0.8J/k_B$ and
$\chi=3.5$ at $T=0.7J/k_B$. These values of $\chi$ are obtained from
measurements of the helicity modulus at the respective temperatures,
and agree with previous studies\cite{simkin}.
For $T=0.8J/k_B$, the $I$-$V$ data at $L_y=512$ is well-described
by the AHNS theory. On the other hand, the envelop of
data sets at smaller values of $L_y$ follows the Minnhagen
{\it et al.} formula (\ref{a_mwjo}). For $T=0.7J/k_B$, we are only able to
obtain reliable data for $I\geq 0.1I_c$ due to the rapid
decrease of $V$ with decreasing $I$. Nevertheless, better agreement
with the AHNS prediction is apparent on the low current side for
the $L_y=512$ array. The larger apparent slope of the data at higher
currents may be attributed to the larger value of $J_{\rm eff}$
due to the smaller length scale probed, which yields a
higher $\chi_{\rm eff}=\pi J_{\rm eff}/k_BT$.

\begin{figure}
\narrowtext
\epsfxsize=\linewidth
\epsffile{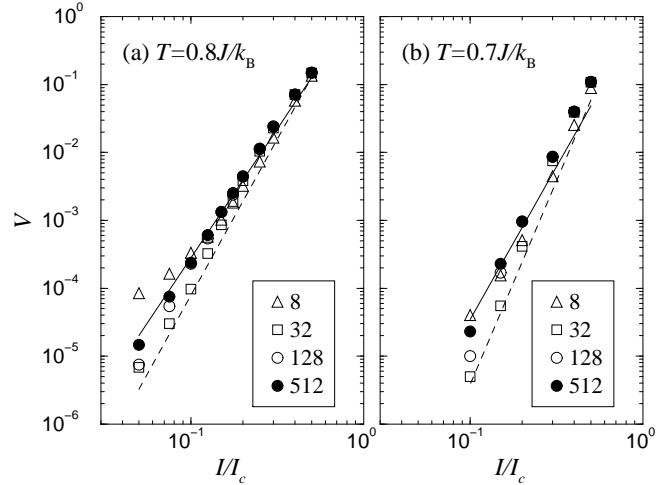}
\caption{Current-voltage curves for arrays of various strip width $L_y$
at two temperatures below $T_{\rm KT}$: 
(a) $T=0.8J/k_B$, and (b) $T=0.7J/k_B$. The value of $L_y$ for
each data set is given in the legend box.
Also shown are the predicted power-laws by AHNS (solid line) and
by Minnhagen \it et al.\rm (dashed line) at each temperature.
}
\label{fig1}
\end{figure}

\begin{figure}
\narrowtext
\epsfxsize=\linewidth
\epsffile{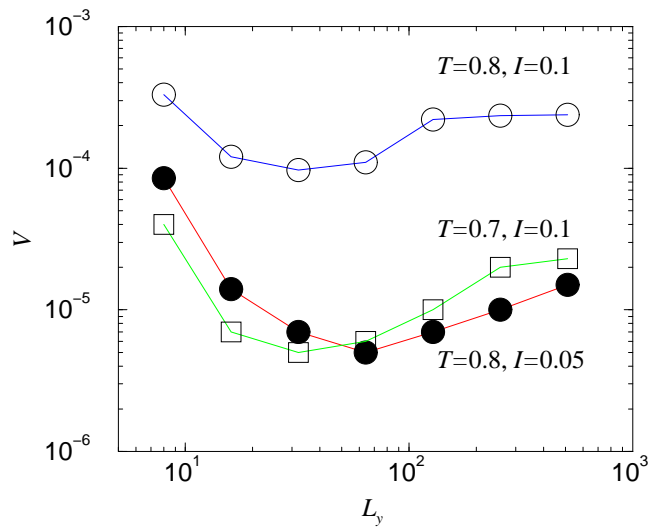}
\caption{Nonmonotonic finite-size dependence of the voltage on the
strip width for three different sets of current and temperature
values.
}
\label{fig2}
\end{figure}

Figure 2 shows three sets of voltage data against transverse array 
size $L_y$. Data at $T=0.8J/k_B$ are represented by open ($I=0.1I_c$)
and solid circles ($I=0.05I_c$), while the set at $T=0.7J/k_B$ and
$I=0.1I_c$ are presented by open squares.
Statistical error of each data point is smaller or comparable to
the symbol size. As $I$ decreases, the size $l_{\rm m}$ where
the minimum value $V=V_m$ is reached increases. $l_{\rm m}$ also varies with
$T$, though this dependence may be weak. Since the change of $V$
with $L_y$ is slow in the neighborhood of $l_{\rm m}$, it may give the false
impression that the finite-size effect is already saturated at $l_{\rm m}$.
As our data shows, the true asymptopic value of $V$ is reached
only at much larger values of $L_y$. At $T=0.7J/k_B$ and $I=0.1I_c$,
for example, the asymptopic value is larger than $V_m$ by five-fold.

In the remaining part of the paper, we propose an explanation
of the unusual finite-size effect, based on the standard picture
of creation and annhilation of free vortices under an applied 
current $I$\cite{blatt}. 
For small $I$, the supercurrent through the network
is maintained by an average phase jump $\Delta\phi=I/I_c$ per column.
Passage of a free vortex across the strip (i.e., in the $y$-direction)
decreases the overall phase advance along the strip by an amount
$2\pi$. In a steady-state situation, this loss is compensated by
a difference in the phase velocities at the two opposing ends of
the strip, and hence a voltage drop across the system. 
Simple counting then yields the following expression for the
average voltage drop per column in a strip of length $L_x\gg 1$
(see, e.g. Ref.\cite{simkin}),
\begin{equation}
V={2\pi\over L_xL_y}\bigl(N_+\bar{v}_+-N_-\bar{v}_-\bigr).
\label{vortex_count}
\end{equation}
Here $N_\pm$ are the number of $\pm$ vortices in the system, and
$\overline{v}_\pm$ are their respective mean velocities 
in the $y$-direction. Although Eq. (\ref{vortex_count}) does
not discriminate between bound and free vortices, it is clear that
correlated drift of a vortex-antivortex pair does not contribute to $V$.

In the AHNS theory, the energy of a vortex/antivortex pair of size $R$
and oriented perpendicular to the external current is given by,
\begin{equation}
E(R)=2E_c+2\pi J\Bigl(\ln R-{I\over I_c}R\Bigr),
\label{pair_energy}
\end{equation}
where $E_c$ is the core energy of a single vortex. 
Equation (\ref{pair_energy}) has a maximum 
$E=E_B=2E_c+2\pi J\ln(I_c/I)-2\pi J$ at
$R=l_{\rm b}=I_c/I$. This is the activation energy for pair breaking.
The rate for such processes follows the Arrhenius law,
\begin{equation}
\Gamma \sim e^{-E_B/k_BT}\simeq (I/I_c)^{2\chi}.
\label{creation_rate}
\end{equation}
In the second step of the argument, one writes down a mean-field 
rate equation governing the density $\rho=\rho_\pm$ of free vortices,
\begin{equation}
\dot{\rho}=\Gamma-\gamma\rho^2,
\label{mf_eq}
\end{equation}
where $\gamma$ is the diffusion constant. The steady-state
vortex density is then given by $\rho_\infty=(\Gamma/\gamma)^{1/2}$.
Going back to Eq. (\ref{vortex_count}), one finds
\begin{equation}
V_\infty\sim \rho_\infty\bar{v}_\pm\sim I^{\chi+1},
\label{V_infty}
\end{equation}
where we have assumed the drift
velocity $\bar{v}_\pm$ of a free vortex to be proportional to $I$.

In the RSJ dynamics, the spin-waves have a 
wavelength-independent relaxation time, of the order of the basic time
unit $\sigma\hbar/2eI_c$. Hence the equilibrium interaction between
vortices in (\ref{pair_energy}) is un-retarded in the dynamic case. 
Equation (\ref{mf_eq}), on the other hand, is based on the assumption that
there is no spatial correlation between vortices and antivortices.
It turns out that this assumption is violated when the width
of the strip is less than a certain characteristic size
$l_{\rm r}=\rho_\infty^{-1/2}\sim (I_c/I)^{\chi/2}$. 
In a bulk system, there is typically one free vortex/antivortex
pair in an area of linear size $l_{\rm r}$ at
any given time. This is no longer the case when $L_y<l_{\rm r}$.
Under periodic boundary conditions in the $y$-direction,
a vortex may annhilate with an antivortex created from the same
bound pair. Consequently, the life time of the two are shortened, 
giving rise to a spurious lower voltage as seen in Fig. 2.

To obtain a semi-quantitative estimate of the voltage reduction,
let us consider a simplified model of vortex-antivortex recombination. 
In a coarse-grained description, we take $l_{\rm b}$ to be the basic
unit of length. The time for a single vortex to traverse
the width of the strip is $\tau=L_y/\bar{v}_\pm=L_yl_{\rm b}$. 
During this period, the typical displacement of the vortex
in the $x$-direction is $l=\tau^{1/2}$.
The probability for the vortex to be within a distance $l_{\rm b}$ from
its initial position after one round is thus 
$p_1=l_{\rm b}/l=(l_{\rm b}/L_y)^{1/2}$.
This is the probability for a vortex-antivortex pair to recombine
in one round across the strip. Assuming $p_1\ll 1$,
the probability for this to happen in two rounds is 
$p_2\simeq (l_{\rm b}/2L_y)^{1/2}$. In general, $p_n\simeq p_1n^{-1/2}$
provided $p_1n^{1/2}<1$. The survival probability after $n$ rounds is 
$1-\sum_{i=1}^np_i\simeq 1-p_1n^{1/2}$. 
This yields a self-recombination time, 
\begin{equation}
\tau_s\simeq p_1^{-2}\tau=L_y^2.
\label{self-rec-time}
\end{equation}
The density of free vortices in this case
is $\rho_s\simeq\Gamma\tau_s\sim L_y^2(I/I_c)^{2\chi}$.
From Eq. (\ref{vortex_count}) we obtain,
\begin{equation}
V_{\rm s}\sim \rho_s \bar{v}_\pm\sim L_y^2(I/I_c)^{2\chi+1},
\label{V_inter}
\end{equation}
which increases with increasing $L_y$.

The typical displacement of the vortex in the $x$-direction over $\tau_s$ 
is $\tau_s^{1/2}=L_y$. To preempt self-recombination, 
we need $\Gamma L_y^2\tau_s\geq 1$, i.e., creation of
the second vortex/antivortex pair in the same region in
space within the time $\tau_s$. This yields
$L_y=(I_c/I)^{\chi/2}=l_{\rm r}$!
Therefore Eq. (\ref{V_inter}) applies in the intermediate
regime $l_{\rm b}\ll L_y\ll l_{\rm r}$. 

The situation is somewhat different when $L_y<l_{\rm b}$. In this case,
the pair-breaking interaction in (\ref{pair_energy}) due to the 
external current is always weaker than the vortex-antivortex attraction
and hence can be treated as a perturbation. The $I$-$V$ curves can
in principle be calculated in a linear response theory\cite{minn87}.
Qualitatively, though, we may estimate the
finite-size effect from the equilibrium relaxation time of a
vortex-antivortex pair on scale $L_y$,
\begin{equation}
\tau_{\rm eq}\simeq L_y^z,
\label{dynamic}
\end{equation}
where $z=2\chi-2$ is the dynamic exponent. We now divide the strip
into square segments of $L_y$ columns each.
The rate of creating a vortex-antivortex pair of size $L_y$
in such a segment is $\tau_{\rm eq}^{-1}$. In thermal equilibrium,
the vortex/antivortex may drift in either direction around the
strip and recombine to generate a $\pm 2\pi$ phase jump. 
This symmetry is broken by
the external current which enhances the $-2\pi$ jump (along the
current direction) by a factor equal to the ratio between the energy 
difference of the two possibilities and $k_BT$.
This leads to the estimate,
\begin{equation}
V_{\rm Ohm}\sim {1\over \tau_{\rm eq}L_y}\Bigl({IJL_y\over I_c k_BT}\Bigr)
\sim (I/I_c)L_y^{-z}.
\label{linear-resp}
\end{equation}
In the Ohmic regime, $V$ decreases rapidly with increasing $L_y$,
as seen in Fig. 2.

At $L_y=l_{\rm b}$, the two expressions (\ref{V_inter}) and
(\ref{linear-resp}) coincide to give
the minimum value $V_m=(I/I_c)^{1+z}$. This is indeed the
power-law form obtained by Minnhagen {\it et al.} by assuming
$l_{\rm b}$ as the only relevant length scale. 

A direct consequence of the self-recombination picture proposed
above is the intermittent phase difference jumps across a square
array of size less than $l_{\rm r}$.
In the steady state, the phase difference $\Delta\phi$ across the system,
averaged in the direction perpendicular to the current,
is expected to grow linearly with time. However, if there is
only at most one pair of free vortex/antivortex present in the system
at any given time, the growth of $\Delta\phi$ is intermittent.
This is indeed the case for $L_y<l_{\rm r}$, where
the life time of a free vortex/antivortex pair is less than the time
it takes to create the pair. The step-wise growth of $\Delta\phi$
has been observed in a previous numerical study using a square 
array\cite{simkin} which we have confirmed.
A similar phenomenon occurs in the $L_y<l_{\rm b}$ regime, 
although in this case there are occasionally reverse steps.
For a long strip, however, $\Delta\phi$
is a much smoother function of time due to averaging along
the strip. 

In summary, extensive simulations of the Josephson junction array
in strip geometry reveal the existence of three size regimes
with distinct current-voltage relationships.
Arrays of size less than $l_{\rm b}\sim I^{-1}$ are in the Ohmic
regime and obey dynamic scaling as proposed by Minnhagen {\it et al.}
Between $l_{\rm b}$ and a second scale $l_{\rm r}\sim I^{-\chi/2}$,
a reversed size-effect is observed.
The power-law $I$-$V$ dependence of AHNS is shown to hold in the
asymptotic regime. Our study reconciles a long-standing dispute between 
two theoretical approaches, and provides a framework for systematic analysis
of finite-size effects in 2D superfluid and superconducting systems
with equal number of vortices and antivortices.

This work is supported by the Research Grant Council (RGC) of the
Hong Kong SAR under grant HKBU2061, by the Hong Kong
Baptist University under grant FRG/98-99/II-74, and by the National
Science Foundation under Grant No. PHY99-07949.
One of us (QHC) is 
supported in part by the National Nature Science Foundation of 
China under Grant No. 10075039.

\end{multicols}

\end{document}